\documentclass[journal]{IEEEtai}

\usepackage[colorlinks,urlcolor=blue,linkcolor=blue,citecolor=blue]{hyperref}

\usepackage{amssymb}
\usepackage{pifont}
\usepackage{makecell}

\usepackage[pdftex]{graphicx}
\usepackage{color,array}

\usepackage{graphicx}
\usepackage{cite}
\usepackage{amsmath,amssymb,amsfonts}
\usepackage{textcomp}
\usepackage{subfigure}
\usepackage[utf8]{inputenc}

\usepackage[ruled,vlined, norelsize]{algorithm2e}
\usepackage{booktabs}
\usepackage{pifont}
\usepackage[english]{babel}
\usepackage{multicol}
\usepackage{placeins}
\usepackage{amsmath}
\def\BibTeX{{\rm B\kern-.05em{\sc i\kern-.025em b}\kern-.08em
    T\kern-.1667em\lower.7ex\hbox{E}\kern-.125emX}}
  
\usepackage{tabu}
\usepackage{cleveref}

\begin{document}


\title{Deep Reinforcement Learning based Evasion Generative Adversarial Network for Botnet Detection}

\author{Rizwan Hamid Randhawa,
        Nauman Aslam,
        Mohammad Alauthman,
        Muhammad Khalid, 
        Husnain Rafiq


\thanks{Rizwan Hamid Randhawa, Nauman Aslam and Husnain Rafiq are with the Department of Computer and Information Sciences, Northumbria University, Newcastle upon Tyne, UK (e-mail: {rizwan.randhawa, nauman.aslam, husnain.rafiq}@northumbria.ac.uk )}

\thanks{Mohammad Alauthman is with the Department of Information Security, University of Petra, Amman 11196, Jordan (e-mail: mohammad.alauthman@uop.edu.jo)}

\thanks{Muhamamd Khalid is with the Department of Computer Science and Technology, University of Hull, UK (e-mail: m.khalid@hull.ac.uk
)}

}



\maketitle

\begin{abstract}
Botnet detectors based on machine learning are potential targets for adversarial evasion attacks. Several research works employ adversarial training with samples generated from generative adversarial nets (GANs) to make the botnet detectors adept at recognising adversarial evasions. However, the synthetic evasions may not follow the original semantics of the input samples. This paper proposes a novel GAN model leveraged with deep reinforcement learning (DRL) to explore semantic aware samples and simultaneously harden its detection. A DRL agent is used to attack the discriminator of the GAN that acts as a botnet detector. The discriminator is trained on the crafted perturbations by the agent during the GAN training, which helps the GAN generator converge earlier than the case without DRL. We name this model RELEVAGAN, i.e. ["relive a GAN" or deep REinforcement Learning-based Evasion Generative Adversarial Network] because, with the help of DRL, it minimises the GAN's job by letting its generator explore the evasion samples within the semantic limits. During the GAN training, the attacks are conducted to adjust the discriminator weights for learning crafted perturbations by the agent. RELEVAGAN does not require adversarial training for the ML classifiers since it can act as an adversarial semantic-aware botnet detection model. Code will be available at https://github.com/rhr407/RELEVAGAN.
\end{abstract}



\section{Introduction}
\label{sec: intro}

Artificial Intelligence (AI) in cybersecurity has become the new normal. However, AI will take some time to win public trust due to inherent biases and adversarial threats ranging from insiders to intrusion and malware. AI-based models can be biased towards the majority class of data on which they are trained due to the imbalance in datasets. Anomaly samples in publicly available datasets are usually scarce compared to the normal data in low data regimes like cybersecurity. Researchers have used data generation techniques like emulation, synthetic oversampling and generative models to mitigate data biasing.
Similarly, adversarial learning has been a topic of pivotal interest to the research communities for the last decade. A plethora of seminal works have been created to deal with adversarial attacks like poisoning, evasion and transferability \cite{mccarthy2022functionality, papernot2016distillation, randhawa2021security}. Adversarial defence strategies can be based on preprocessing, adversarial training, architecture, detection, defensive testing, multiclassifiers and game theory \cite{mccarthy2022functionality}. 

Adversarial training seems to be a simplistic strategy to provide robustness against adversarial evasion attacks; however, it has some cons. First, increasing the number of samples in the auxiliary data may not have a linear relationship with detection accuracy \cite{randhawa2021security}. Second, the adversarial training can not guarantee a robust defence as it can be bypassed \cite{tramer2017ensemble}. Although adopted as an immediate remedy against some adversarial attacks, it can not be considered an ultimate cure for the grave problem in AI. Third, it consumes additional time for retraining. Several GAN-based research works preserve the functionality of the generated samples by manipulating only non-functional features \cite{usama2019generative, lin2022idsgan, duy2021digfupas}. So, GANs do not play a role in generating a complete feature vector in those works. It is also quite challenging to generate categorical features using a GAN without manual engineering except using a sequence GAN \cite{cheng2021packet}. Researchers have also used deep reinforcement learning to generate the functionality preserving adversarial evasion attacks \cite{apruzzese2020deep, anderson2018learning, wu2019evading}. The main goal of employing deep reinforcement learning is to explore functionality preserving adversarial samples since the DRL can guarantee semantic awareness in contrast to GANs; however, these works consider the adversarial training to make the detection models adept at evasion awareness. 

RELEVAGAN is an effort toward a unifying model concept that would solve the problem of data imbalance, provide adversarial semantic awareness and save training time. RELEVAGAN is equipped with an integrated DRL agent to achieve the said goals. RELEVAGAN name was chosen for two reasons. First, it is a deep REinforcement Learning-based Evasion Generative Adversarial Network. Second, it relieves the employed GAN model to make its job easier by letting the generator of the GAN explore the semantic aware samples, i.e. within certain boundary conditions. We can call this RELiEVe A GAN as well. Either way, RELEVAGAN proves to be an improved technique compared to the peer models like Auxiliary Classifier GAN (ACGAN) and Evasion Generative Adversarial Network (EVAGAN).

The DRL agent attacks the RELEVAGAN's discriminator, which acts as a botnet detector. The attack generation is based on manipulating the real attack samples to evade the botnet detector. As the RELEVAGAN training proceeds, the agent learns to evade the botnet detector. The discriminator is adversarially trained on the evaded samples from the DRL attacker and synthetic samples from the generator in each training iteration. After a certain number of epochs, the discriminator becomes hardened against the samples from the DRL agent and the generator. The detection estimations for the benign, real and generated samples and the generator training settle to the desired values in fewer training iterations than the EVAGAN model. The experimental analysis shows the considerable performance of the RELEVAGAN model against EVAGAN in terms of detection estimation and stability of training for three different botnet datasets. We argue that the learning of GAN follows semantic awareness because GAN is also trained on the attacks generated by the DRL model.

Following are the main contributions of this paper: 

\begin{enumerate}

\item We propose a novel DRL-based GAN model to address the problems of data imbalance, evasion awareness and functionality-preservation in synthetic botnet traffic generation.
\item We demonstrate by experiments that DRL plays a role in GAN training for the detection of synthetic as well as real attacks. 
\item Because of the integrated self-learning DRL attacker, the proposed model can be envisaged as sustainable against evolving botnet.
\item We determine that RELEVAGAN outperforms EVAGAN in terms of early convergence of the training.
\end{enumerate}


\section{Background} 
\label{sec: background}

\begin{table}[tb!]

\centering
\caption{Main notations}

\label{table: main notations}

   \begin{tabular}{lcc}

      \hline
      
      \multicolumn{1}{|c}{\textbf{Notation}}&
      \multicolumn{1}{|c|}{\textbf{Definition}}\\
      
      \hline
      
      \multicolumn{1}{|c}{$\mathcal{G}$}&
      \multicolumn{1}{|c|}{Generator}\\
      
      \hline
      
      \multicolumn{1}{|c}{$\mathcal{D}$}&
      \multicolumn{1}{|c|}{Discriminator}\\
      
      \hline
      
      \multicolumn{1}{|c}{z}&
      \multicolumn{1}{|c|}{Normal distribution from noise space}\\
      
      \hline
      
      \multicolumn{1}{|c}{${z}$}&
      \multicolumn{1}{|c|}{Noise samples}\\
      
      \hline
      
      \multicolumn{1}{|c}{$p_{data}$}&
      \multicolumn{1}{|c|}{Probability distribution of real samples}\\
      
      \hline
      
      \multicolumn{1}{|c}{$p_{z}$}&
      \multicolumn{1}{|c|}{Probability distribution of noise samples}\\
      
      \hline
      
      \multicolumn{1}{|c}{$\mathcal{X}$}&
      \multicolumn{1}{|c|}{Real data distribution}\\
      
      \hline
      
      \multicolumn{1}{|c}{$\mathbb{E}$}&
      \multicolumn{1}{|c|}{Expected value}\\
      \hline

      \multicolumn{1}{|c}{$c_m$}&
      \multicolumn{1}{|c|}{Minority class labels}\\
      \hline
      
      \multicolumn{1}{|c}{$c_M$}&
      \multicolumn{1}{|c|}{Majority class labels}\\
      \hline

      \multicolumn{1}{|c}{$y_{x_{i}}$}&
      \multicolumn{1}{|c|}{Actual label of sample $x_i$ in dataset $\mathcal{X}$}\\
      \hline

   \end{tabular}

 \end{table}

\subsection{Data Imbalance}

Botnet datasets suffer from the imbalance problem. That said, in most of the publicly available network traffic datasets, the number of botnet class samples is meagre compared to the normal traffic samples. This inequitable distribution of data makes the ML-based detection models less accurate. To address this issue, data undersampling can be adopted. However, it can result in the loss of diversity and representation of normal traffic \cite{alfaiz2022enhanced}. Oversampling can also solve the data imbalance problem to some extent; however, the use of nearest neighbours, and linear interpolation, may not be suitable for the high-dimensional and complex probability distributions \cite{chawla2002smote, engelmann2020conditional}. Researchers have tested several oversampling techniques in \cite{kovacs2019empirical} to rank the best performing being SMOTE\_IPF, ProWSyn and polynom\_fit\_SMOTE. However, authors in \cite{randhawa2021security} declared GANs outperforming those three oversamplers in most of the adversarial training of ML classifiers. Hence we can consider GANs as a suitable candidate for data oversampling compared to other synthetic data generation methods.

\subsection{Generative Adversarial Nets}
\label{subsec: gans}
A GAN is a combination of two neural networks having different structures: generator and discriminator.  The generator ($\mathcal{G}$) synthesises samples, and the discriminator ($\mathcal{D}$) evaluates those samples. The GAN training is completed in two steps. First, the $\mathcal{D}$ is trained on real data with the label REAL, and the data synthesised by an untrained $\mathcal{G}$ with the label FAKE. Next, the trained $\mathcal{D}$ in the previous step is tested on the fake/synthetic data coming from $\mathcal{G}$ labelled as REAL. $\mathcal{G}$ updates its weights based on the loss output coming from $\mathcal{D}$ on this falsely labelled data in a batch training. Several batch iterations make one complete traversal of the dataset called an epoch. Mathematically, $\mathcal{G}$ can be represented as $\mathcal{G}$: z$\,\to\,$ $\mathcal{X}$ where z can be the normal or uniform distribution and $\mathcal{X}$ represents the real data distribution. 

In a classical GAN, $\mathcal{D}$: $\mathcal{X}$ $\,\to\,$[0,1] acts as a classifier to give us an estimate of probability (between 0 and 1) to mark whether the input data is real or fake. The objective function of the combined model is denoted by Equation \ref{eq1}.

\begin{equation} \label{eq1}
\begin{aligned}
\min_\mathcal{G} \max_\mathcal{D} V(\mathcal{\mathcal{D}}, \mathcal{G})= \mathbb{E}_{x\sim p_{data}(x)}[\log \mathcal{D}(x)] + \\ \mathbb{E}_{z\sim p_z(z)}[\log(1 - \mathcal{D}(\mathcal{G}(z)))]
 \end{aligned}
\end{equation}

In Equation \ref{eq1}, $\mathbb{E}$ denotes the expected value of the loss, and $x$ and $z$ represent the real and noise data samples consecutively. Similarly, $p_{data}$ and $p_{z}$  are the probability distributions of real data and noise, respectively. The min-max game minimises the $\mathcal{G}$'s loss in generating data similar to the real data. Since $\mathcal{G}$ is not able to control $\mathcal{D}$'s loss on real data, it can maximise the loss of $\mathcal{D}$ on generated data $\mathcal{G}(z)$. The objective function of $\mathcal{G}$ is denoted by Equation \ref{eq2}.

\begin{equation} \label{eq2}
\begin{aligned}
J^\mathcal{G}(\mathcal{G})= \mathbb{E}_{z\sim p_z(z)}[\log(\mathcal{D}(\mathcal{G}(z)))]
 \end{aligned}
\end{equation}

One complete iteration of the GAN training takes noise as input and the output of the $\mathcal{D}$ as the feedback to update the weights of the $\mathcal{G}$. This training process keeps iterating for a specific number, after which $\mathcal{G}$ and $\mathcal{D}$ do not learn further, resulting in Nash equilibrium. 

\subsection{Evasion Awareness}

Evasion awareness is defined as the ability of a detection model to foresee the manipulation of the input samples to intentionally fool it by an adversary. The inherent problem of decision bias in ML classifiers can lead to evasion attacks, especially in low data regimes. An adversarial evasion j* is a crafted version of an input sample j given by Equation \ref{eq3}, where $\eta$ is a perturbation, and $\epsilon$ is the accepted range. When making an adversarial attack, $\eta$ is sought and selected so that the classifier can not differentiate between j* and j \cite{papernot2017practical, apruzzese2020deep, wu2019evading}.

\begin{equation} \label{eq3}
    \begin{aligned}
    j^{*} = |j + \eta| \le \epsilon
    \end{aligned}
\end{equation}

Adversarial training is widely adopted to proactively make the ML classifiers aware of the evasion samples. However, extending the capability of $\mathcal{D}$ of a GAN from discriminating between real and fake samples to differentiating between normal and anomaly data renders the adversarial training needless. Hence $\mathcal{D}$ can be used as an evasion-aware classifier \cite{yin2018enhanc, yin2019enhancing}. In [evagan], authors propose EVAGAN that provides such type of $\mathcal{D}$ and compare its performance with the $\mathcal{D}$ of ACGAN and other ML classifiers like xgboost (XGB), naive bayes (NB), decision tree (DT), random forests (RF), k-nearest neighbours (KNN) and logistic regression (LR). EVAGAN's $\mathcal{D}$ outperforms the ML classifiers in black box testing and gives 100\% accuracy in normal and evasion samples estimation. However, EVAGAN, like other GAN models, is agnostic of semantics and functionality preservation of malicious synthetic samples. In this paper, we propose a novel type of GAN based on EVAGAN leveraged with deep reinforcement learning to address the problem of functionality preservation. 

\subsection{ACGAN}
The proposed RELEVAGAN is based on EVAGAN, which is an improved version of ACGAN. ACGAN's $\mathcal{D}$, along with differentiating between real or fake samples, also considers the class labels in the training process \cite{odena2017conditional}. That said, $\mathcal{D}$ of ACGAN works as a dual classifier for differentiating between the real/fake data and different classes of the input samples, whether coming from the real source or the $\mathcal{G}$. In addition to random noise samples $z$, the input of $\mathcal{G}$ in ACGAN includes class labels $c$. The generator of the ACGAN can generate the specified class data for which the labels are fed to its $\mathcal{G}$. The objective function of ACGAN can be derived using likelihoods denoted by $L_S$ and $L_C$ for the correct source data and real class labels, respectively. $\mathcal{D}$ trains to maximise $L_{C} + L_{S}$ and $\mathcal{G}$ learns to maximise $L_{C} - L_{S}$. Hence, the objective of $\mathcal{D}$ is to improve the two likelihoods, while the aim of $\mathcal{G}$ is to assist $\mathcal{D}$ in improving the performance on class labels. $\mathcal{G}$'s other target is to suppress the log-likelihood of $\mathcal{D}$ on fake samples. The $\mathcal{D}$ outputs probability distributions over sources and the class labels [$P (S | \mathcal{X}), P (C | \mathcal{X})] = \mathcal{D}(\mathcal{X})$ where $S$ denotes the sources (real/fake) and $C$ represents the class labels. Equations \ref{eq3} and \ref{eq4} represent the $L_s$ and $L_c$ respectively.

\begin{equation} \label{eq4}
\begin{aligned}
L_{S} = \mathbb{E}[log P (S = real | \mathcal{X}_{real} )]\ + \\
\mathbb{E}[log P (S = fake | \mathcal{X}_{fake} )]
 \end{aligned}
\end{equation}

\begin{equation} \label{eq5}
\begin{aligned}
L_{C} = \mathbb{E}[log P (C = c| \mathcal{X}_{real} )]\ + \\
\mathbb{E}[log P (C = c | \mathcal{X}_{fake} )]
 \end{aligned}
\end{equation}

\begin{figure}[tb!]
\centering
\includegraphics[width=8.8cm, height=5cm]{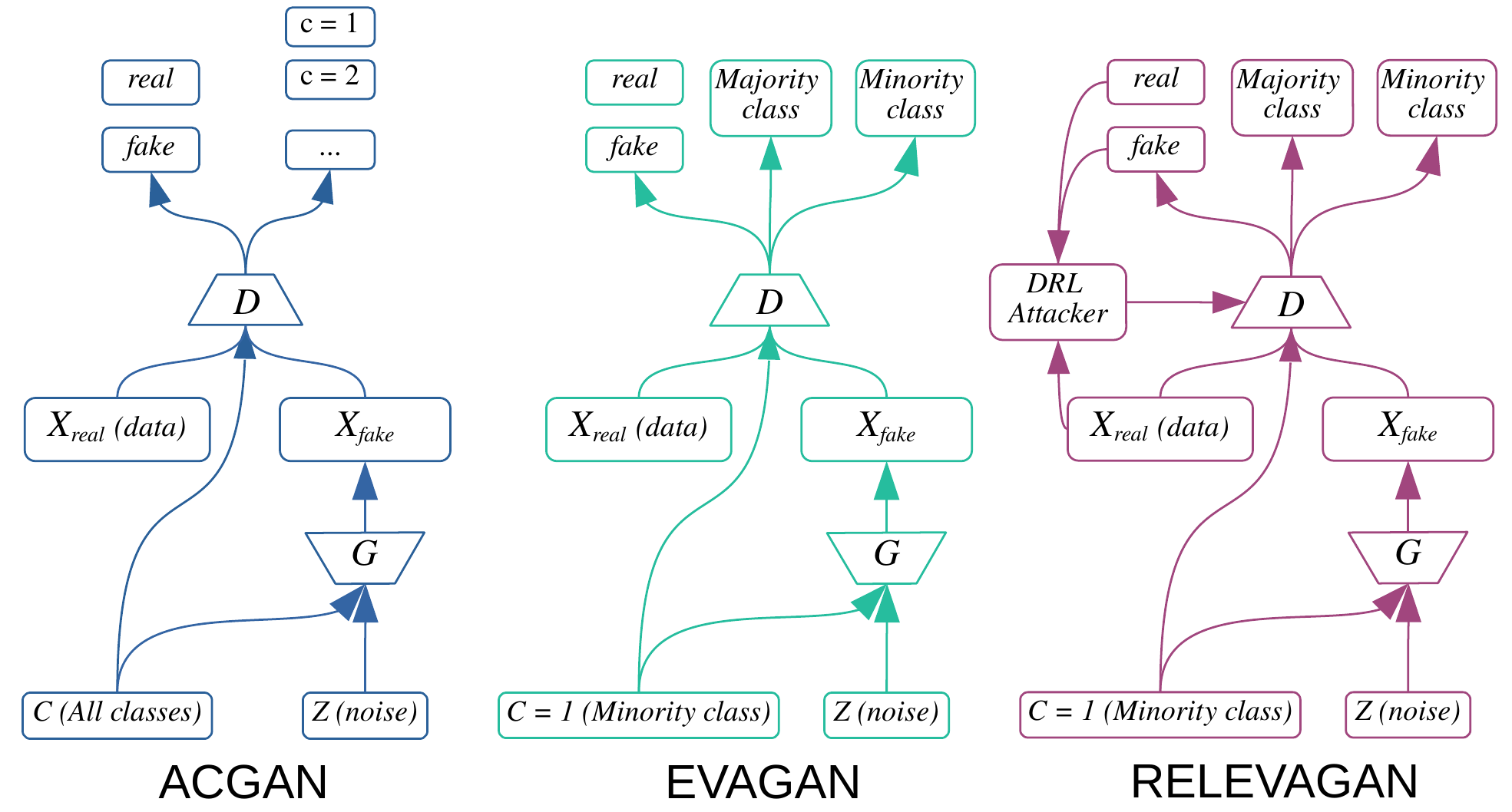}
\caption{Comparison of RELEVAGAN model with ACGAN and EVAGAN}
\label{fig: ACGAN_vs_EVAGAN_vs_RELEVAGAN}
\end{figure}

\subsection{EVAGAN}
\label{subsec: evagan}

EVAGAN is a specialised version of ACGAN with binary class consideration for low data regimes. The $\mathcal{G}$ of ACGAN considers generating multiple class samples $\mathcal{X}_{fake} = \mathcal{G} (c, z)$ where $c$ is the class label. Hence the number of the samples can include $C = \{c_1, c_2, c_3, ..., c_n\}$ which may not be required in low data regimes as we need to generate only the anomaly samples which are low in number. The generator does not need to improve the classification performance of $\mathcal{D}$ on normal/majority class samples. This considerably reduces the training time of $\mathcal{G}$ as the diversity of the input samples from a single class is less intricate. In this way, we can improve the performance of the $\mathcal{G}$ and harden the $\mathcal{D}$ simultaneously with fewer $c_m$ samples. EVAGAN also hardens itself on the evasion samples generated by its $\mathcal{G}$ which takes noise $n$ and the single class labels $c$ = 1. The minority class labels are embedded into the input layer of the $\mathcal{G}$. Since the $\mathcal{G}$ only needs to generate minority class samples so it should only consider the loss of $\mathcal{D}$ on the estimation of minority class and the sources. The goal of $\mathcal{G}$ is to maximise the $\mathcal{D}$'s loss on the fake source and assist in reducing the $\mathcal{D}$'s loss on minority class samples. Let $y_{x_{i}}$ denotes the actual label of sample $x_i$ in dataset $\mathcal{X}$, $P(S = fake | \mathcal{X}_{m_{fake}})$ represent the prediction of probability distribution of samples being fake and $P(C = c_m | \mathcal{X}_{m_{fake}})$ is the predicted output for probability distribution from $\mathcal{D}$ for minority class labels $c_m$, then the loss function of $\mathcal{G}$ for $N$ samples will be given by the Equation \ref{eq6}.

\begin{equation} \label{eq6}
\begin{aligned}
G\_Loss = - \frac{1}{N}\sum\limits_{i = 1}^{N} [
y^{fake}_{x_{i}} (\log P(S = fake | \mathcal{X}_{m_{fake}})) + \\
y^{c_m}_{x_{i}} (1 - \log P(C = c_m | \mathcal{X}_{m_{fake}})) ]
 \end{aligned}
\end{equation}

In Equation \ref{eq6}, $y^{fake}_{x_{i}}$ and $y^{c_m}_{x_{i}}$ represent the actual labels for fake and minority classes respectively. The primary goal of $\mathcal{G}$ is to minimise the $G\_Loss$, so it inclines towards reducing the correct estimation of $\mathcal{D}$ on fake samples by suppressing the term $\log P(S = fake | \mathcal{X}_{m_{fake}})$. For the second objective, it will try to increase the value of $\log P(C = c_m | \mathcal{X}_{m_{fake}})$ so that the second term in the equation can also be suppressed in value. The loss function of $\mathcal{D}$ is given by Equation \ref{eq7}.

\begin{equation} \label{eq7}
\begin{aligned}
D\_Loss = - \frac{1}{N}\sum\limits_{i = 1}^{N} [
y^{c_M}_{x_{i}} (\log P(S = c_M | \mathcal{X}_{M_{real}})) + \\
y^{real}_{x_{i}} (\log P(S = real | \mathcal{X}_{m_{real}})) + \\
(1 - y^{real}_{x_{i}}) (1 - \log P(S = real | \mathcal{X}_{m_{real}})) + \\
y^{c_{m_{real}}}_{x_{i}} (\log P(C = c_m | \mathcal{X}_{m_{real}})) + \\
(1 - y^{c_{m_{real}}}_{x_{i}}) (1 - \log P(C = c_m | \mathcal{X}_{m_{real}})) ]
 \end{aligned}
\end{equation}

In Equation \ref{eq7}, the loss of $\mathcal{D}$ is calculated as per the three different binary cross-entropy losses for majority class, sources and minority class estimations. 

\subsection{Semantics/Functionality Preservation}
Semantics or functionality preservation means that the synthetic samples can still execute the original malicious function intended by the attacker. Various researchers have employed GANs to synthesise the network traffic data to address the issue of data imbalance, and adversarial evasion attacks \cite{randhawa2021security, huang2020igan, shahriar2020g}. However, GANs are not good at generating categorical features \cite{cheng2021packet}. Researchers have used one-hot representation \cite{usama2019generative}, or IP2Vec techniques \cite{IP2Vec, ring2018flow} to transform IP addresses into integer values to input to GANs for data generation. Since GANs follow the probability distribution of the input data so there is a high chance that the synthetic samples may lie outside the semantic limits. The perturbations could be easily perceived as anomalies by the ML detectors. To preserve the functionality, the perturbation j* in Equation \ref{eq3} should be small enough not to be perceived by the detectors as malicious and distant enough to be considered normal samples. At the same time, the malicious activity must not be compromised. For this reason, several researchers have proposed functionality preservation by modifying only the non-functionality preserving features using GANs \cite{usama2019generative, cheng2021packet, duy2021digfupas}. However, various research works claim to preserve the malicious functionality of the generated samples using deep reinforcement learning.

\subsection{DRL-based Evasion Generation}

A reinforcement learning model comprises an agent and an environment that engage for a defined number of iterations/steps. For a turn $t$ the agent chooses an action $a_{t}$ $\epsilon$ $\mathcal{A}$ using some policy $\pi(a|s_{t} )$ and an observable state $s_{t}$. The environment returns a reward $r_{t}$ $\epsilon$ $\mathbb{R}$ against an action and a new state $s_{t+1}$. The reward $r_{t}$ and observed state of the environment $s_{t+1}$ are fed back to the agent for defining a new action based on policy $\pi(a|s_{t+1} )$. Keeping the current state in view, the agent tries to learn through the trade-off between exploration and exploitation. The maximum reward collection is the main goal for exploration, which can be possible if the agent employs a certain policy to boost the expected value given by the Q-value function in Equation \ref{eq_V}.

\begin{equation} \label{eq_V}
    \begin{aligned}
    V^{\pi}(s_{t}) = \mathbb{E}_{a_{t}} [Q^{\pi}(s_{t}, a_{t})| s_{t}]
    \end{aligned}
\end{equation}

Here $Q$ is given by Equation \ref{eq_Q}.

\begin{equation} \label{eq_Q}
    \begin{aligned}
    Q^{\pi}(s_{t}, a_{t}) = \mathbb{E}_{s_{t+1: \infty}, a_{t+1: \infty}}[R_{t}|s_{t}, a_{t}] 
    \end{aligned}
\end{equation}

and $R$ is denoted by Equation \ref{eq_R}

\begin{equation} \label{eq_R}
    \begin{aligned}
    R_{t} = \Sigma_{i\geq 0} \gamma^i r_{t+i} 
    \end{aligned}
\end{equation}

In Equation \ref{eq_R}, $\gamma \in [0, 1]$ is a discounting factor for rewards from future actions. For catering to the large intractable storage of tabular representation of the Q-function, deep neural networks come to the rescue to act as an approximator in the deep Q-Network (DQN) to represent the state-action value function \cite{mnih2015human}. In RELEVAGAN, we use double DQN as the core technique for DRL-based attacks generation.

\begin{figure}[tb!]
\centering
\includegraphics[width=6cm, height=6cm]{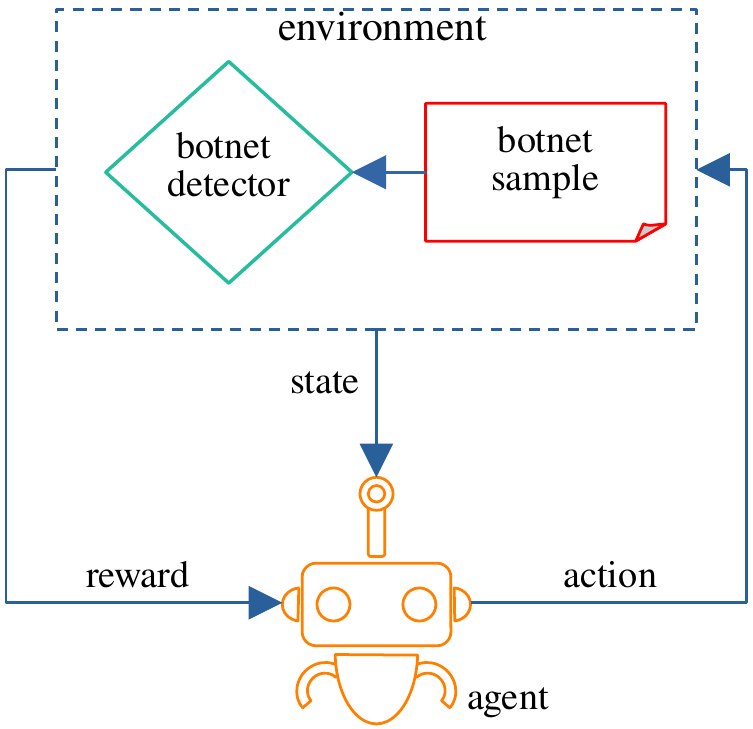}
\caption{Markov Decision Process for Botnet evasion generation using Reinforcement Learning}
\label{fig: DRL}
\end{figure}

Figure \ref{fig: DRL} shows the Markov Decision Process (MDP) \cite{alauthman2016efficient} where an agent is interacting with a botnet detector environment by providing a new sample which can act as a potential evasion. The Q-function and the policy help in taking a particular action. The botnet sample is a manipulated feature vector defined within certain boundaries to ensure semantics preservation. The reward is the classification output of the botnet detector. The reward and the new state are fed back to the agent after trying the botnet detector.

\subsection{DRL-based Functionality Preservation}

To preserve the malicious functionality, researchers generate the evasion samples within semantic limits \cite{wu2019evading, mao2022evaderl, anderson2018learning, fang2019evading, apruzzese2020deep } by attacking the trained classifiers using a DRL agent. The evaded samples were collected to be used for adversarial training to make the model adept at adversarial awareness. Similarly, in \cite{mao2022evaderl}, authors used DRL to attack PDF (Portable Document Format) malware detectors to generate evasive malware samples keeping the functionality of the attacks preserved. The rationale for using the DRL is that we can bound it to explore the evasion attacks within a specific range defined by $\epsilon$ in Equation \ref{eq3}. In this paper, we have used the DRL and the EVAGAN model to continuously generate the evasion attack samples using a DRL agent. The process becomes part of the EVAGAN training. The marriage of GAN and DRL culminates RELEVAGAN, which uses the power of EVAGAN as a robust evasion-aware detection model and DRL as the semantics check on the samples generated by the $\mathcal{G}$ of EVAGAN. The details of the model will be further discussed in Section \ref{sec: relevagan}.

\begin{figure*}[tb!]
\centering
\includegraphics[width=\textwidth, height=9.5cm]{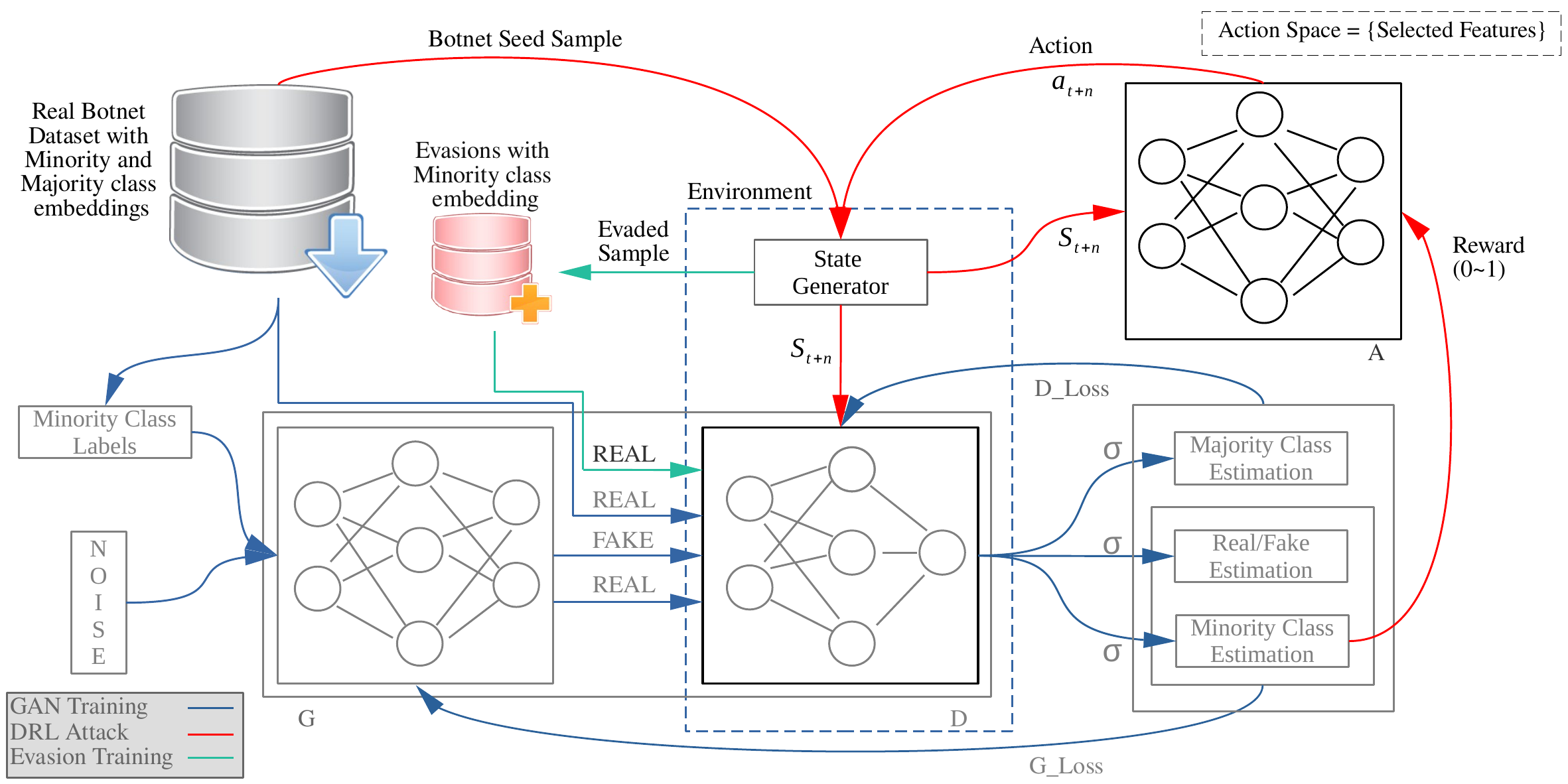}
\caption{RELEVAGAN Architecture}
\label{fig: RELEVAGAN}
\end{figure*}

\section{RELEVAGAN}
\label{sec: relevagan}

In this section, we discuss the motivation behind the design of RELEVAGAN, especially the structural explanation of its DRL attacker. As illustrated by Figure \ref{fig: ACGAN_vs_EVAGAN_vs_RELEVAGAN}, the DRL attacker has been introduced in RELEVAGAN. The rest of the architecture is similar to EVAGAN. The impact of using the DRL attacker helps improve the EVAGAN performance, the details of which will be discussed in section \ref{sec: relevagan}. 
\subsection{Motivation}
Generative adversarial-based evasion samples generation follows the probability distribution of the input data samples as the $\mathcal{G}$ of a GAN trains itself based on the feedback coming from $\mathcal{D}$. The $\mathcal{G}$ tries to explore the new sample spaces that are unknown by the $\mathcal{D}$ so that it can fool the $\mathcal{D}$; however, this process can lead to the creation of samples that may not follow the real malicious semantics. To address this issue, deep reinforcement learning-based samples can help the $\mathcal{G}$ learn the boundaries of the real samples. For this reason, a DRL-based agent can be leveraged to explore samples in defined observation space. The $\mathcal{D}$ can be trained on these generated samples, which also gives the feedback to the $\mathcal{G}$ in the GAN training. Eventually, the $\mathcal{G}$ can start learning from the updated feedback from $\mathcal{D}$ and generates the samples within a defined range set by the DRL agent. This process can help converge the $\mathcal{G}$ training earlier while at the same time achieving high accuracy in a lesser number of epochs. Although semantic awareness comes with additional training costs for the DRL part but motivated by this rationale, the RELEVAGAN is a step further toward a more intelligent functionality preserving GAN design.

\subsection{Architecture}
Figure \ref{fig: RELEVAGAN} shows the architecture of RELEVAGAN. The structure of RELEVAGAN has mainly two components. One is EVAGAN, summarised in Section \ref{subsec: evagan}, and the other part is a DRL-based model. For more details on EVAGAN, readers are highly encouraged to refer to the paper \cite{randhawa2021evagan}. A typical DRL model consisting of an agent and an environment has been coupled with EVAGAN architecture in RELEVAGAN design. Like other DRL-based attackers for evasion generation using a black box attack \cite{wu2019evading, mao2022evaderl, anderson2018learning, fang2019evading, apruzzese2020deep }, our proposed DRL agent attacks the $\mathcal{D}$ of EVAGAN, which acts as a black box classifier. $\mathcal{D}$'s output for minority class estimation is used as the reward for the agent to adjust its weights and generate a new action $a_{t+n}$ based on some policy $\pi$. The new action is fed to the environment where the state generator creates a new state taking another seed sample from the real data set. As the result of a single iteration, the new state and the collected reward are fed back to the agent. As a result of the positive reward, the evasion samples are fed to the $\mathcal{D}$ of EVAGAN to adversarially train it. In this way, the $\mathcal{D}$ becomes proactively aware of any possible future evasions and ready to give better feedback to the $\mathcal{G}$ to train to confine its boundaries for evasion generation. The process leads to the early convergence of $\mathcal{G}$ training.

\subsection{Environment}
The environment in RELEVAGAN consists of mainly two parts: 
\subsubsection{State Generator}
The state generator is responsible for three different jobs: 
\begin{itemize}
    \item It takes a botnet seed sample as the current state $S_{t}$ from the real botnet dataset and transforms it into a feature vector accepted by the $\mathcal{D}$ of EVAGAN based on some action index $n$ coming from the agent. 
    \item It feeds back the new state  $S_{t+n}$ to the agent.
    \item If the sample is evaded by the $\mathcal{D}$, the state generator is also responsible for storing and/or feeding it to the $\mathcal{D}$ to adversarially train in parallel with EVAGAN training.

\end{itemize}
\subsubsection{Botnet Detector}
The target model is the $\mathcal{D}$ of EVAGAN. In Figure \ref{fig: RELEVAGAN}, the EVAGAN model has been illustrated in grey-coloured border lines, and the difference has been highlighted in black lines for a better understanding of where the RELEVAGAN is different from EVAGAN.
\subsection{Action Space}
\label{subsec: action_table}
The following feature set gives the action space through which the agent chooses the most appropriate index $n$ to gain the maximum reward by evading the target botnet model. These features have been chosen based on the work in paper \cite{apruzzese2020deep} for the three datasets used in this work. The details of the datasets are mentioned in Section \ref{sec: ImplementationDetails}.

\begin{itemize}
    \item FlowDuration
    \item FlowBytes/s 
    \item FlowPackets/s
    \item FwdPackets/s
    \item BwdPackets/s
    \item TotalLengthofFwdPacket
    \item TotalLengthofBwdPacket
    \item BwdPackets/s
    \item SubflowFwdBytes
    \item FwdHeaderLength
    \item BwdHeaderLength
    \item Down/UpRatio
    \item AveragePacketSize

\end{itemize}

To keep the functionality reservation, we limit the change in the feature value to $\Delta$, which is the minimum value of the particular feature within the data set as given by Equation \ref{eq_delta}.

\begin{equation} \label{eq_delta}
    \begin{aligned}
    \Delta_{n} = \min_{\forall m \in F_n} X(m)
    \end{aligned}
\end{equation}

In Equation \ref{eq_delta}, $\Delta_{F_n}$ is the minimum value for all the rows $m$ of a particular feature $F$ and $n$ is the action index coming from the agent as a particular feature number from the action table.
 
\subsection{Agent} 
The agent is a deep neural network with the size of the observation space as input and the number of actions as the output. The observation space in RELEVAGAN is a complete botnet sample as a feature vector. The agent is responsible for choosing an action index $n$ among the features in the action table as mentioned in Subsection \ref{subsec: action_table}. Based on this index, a training step is executed, which feeds back the reward and the new state to the agent. 
\subsection{Reward}
\label{subsec: reward}
The reward in a typical botnet evasion generation model using a DRL black-box attack is the output of the botnet detector \cite{wu2019evading, apruzzese2020deep}, which can be a real number in the range of [0, 1]. In our case, for the botnet sample, the expected value is '0', and for a normal traffic sample, the output should be ideally '1'. Hence for a botnet sample to be considered a successful evasion by a DRL attacker, we set the threshold for the reward to be greater than 0.5. In other terms, if the sample generated by the DRL attacker is evaded with more than 50\% confidence, the reward will be '1' and '0' otherwise.
\subsection{Training}
RELEVAGAN training is similar to EVAGAN except for adding a couple of extra steps for the DRL agent after every batch training. In Algorithm \ref{algo: RELEVAGAN}, steps 3 and 4 discriminate the training between EVAGAN and RELEVAGAN for a defined number of batches. The sequence of the steps is crucial for understanding the rationale behind RELEVAGAN. Note that we train $\mathcal{G}$ after the evasion training of $\mathcal{D}$. Since $\mathcal{D}$'s weights are adjusted as per the evasions generated by the DRL attacker in Step 4, it will feed the $G\_Loss$ back to$\mathcal{G}$ more cognitively as compared to the case of EVAGAN training. The DRL attack is executed in every batch of training. 
\begin{algorithm}
\For{i= 1, 2, 3, ..., number of batches}{ \BlankLine
Step 1: Train $\mathcal{D}$  on real data\\
Step 2: Train $\mathcal{D}$  on generated data\\
\textbf{Step 3}: Execute DRL $\mathcal{A}$for generating evasion on batch size\\
\textbf{Step 4}: Train $\mathcal{D}$  on $\mathcal{A}$  generated evasions\\
Step 5: Train $\mathcal{G}$ \\
}
\caption{RELEVAGAN Training}
\label{algo: RELEVAGAN}
\end{algorithm}

\section{Implementation Details}
\label{sec: ImplementationDetails}
\subsection{Experimental Setup}
As mentioned in the EVAGAN paper, the experiments for the RELEVAGAN were performed on a GPU workstation, AMD Ryzen threadripper 1950x, equipped with a 16-core processor and an 8GB memory GeForce GTC 1070 Ti graphics card. The OS used was ubuntu 20.04, running Keras, TensorFlow, Sklearn and Numpy libraries within the Jupyter notebook. The source code of RELEVAGAN is also available on GitHub under MIT license\footnote{https://www.github.com/rhr407/RELEVAGAN}.

\subsection{Data Preparation}
For experimentation, we have used three different botnet datasets, ISCX-2014, CIC-2017 and CIC-2018, from the Canadian Institute of Cybersecurity (CIC) for the quantitative analysis of RELEVAGAN. The choice of the dataset is based on the work done by authors in \cite{randhawa2021security}. An open-source tool, CICFlowMeter-v4 from CIC, was used for feature extraction. The choice of feature set for training and data preprocessing is the same as used in EVAGAN. Table \ref{table: number of samples in datasets} shows the distribution of benign and botnet samples in all three datasets. The details of a particular botnet selection are mentioned in EVAGAN paper \cite{randhawa2021evagan}. 
\begin{table}[ht!]

\centering
\caption{Distribution of normal and botnet samples in cybersecurity botnet datasets}

\label{table: number of samples in datasets}

   \begin{tabular}{lcccc}

      \hline

      \multicolumn{1}{|c}{\textbf{Dataset}}&
      \multicolumn{1}{|c}{\textbf{Normal}}&
      \multicolumn{1}{|c}{\textbf{Real\_bots}}&
      \multicolumn{1}{|c|}{\textbf{Total Samples}}\\
      
      \hline
      
      \multicolumn{1}{|c}{ISCX-2014}&
      \multicolumn{1}{|c}{246929}&
      \multicolumn{1}{|c}{Virut: 1748}&
      \multicolumn{1}{|c|}{248677}\\
      
      \hline
      
      \multicolumn{1}{|c}{CIC-IDS2017}&
      \multicolumn{1}{|c}{70374}&
      \multicolumn{1}{|c}{Ares: 1956}&
      \multicolumn{1}{|c|}{72330}\\
      
      \hline
      
      \multicolumn{1}{|c}{CIC-IDS2018}&
      \multicolumn{1}{|c}{390961}&
      \multicolumn{1}{|c}{Ares/Zeus: 2560}&
      \multicolumn{1}{|c|}{393521}\\
      
      \hline
    
   \end{tabular}

 \end{table}

\subsection{DRL Attacker}
For the implementation of the DRL attacker, we used the OpenAI Gym, and gym-malware tool kits \cite{brockman2016openai, anderson2018learning}. Keras-rl and Keras-rl2 libraries were used for the selection of the DQN agent. Unlike a typical DRL algorithm, we execute a new training session in every batch of the RELEVAGAN training where the weights of the neural network are not reset to ensure that the agent is learning in each batch iteration of RELEVAGAN. The reason for keeping the weights is that we can not estimate the number of training iterations that would e traverse the whole batch of botnet samples for generating manipulations. Hence, we reinitialise the session after each batch keeping the agent's neural network unchanged. A single training session in each batch iteration lasts until the following two cases appear:
\begin{itemize}
    \item The evasion is successful.
    \item The number of tries saturates.
\end{itemize}
In either of the cases mentioned above, a new botnet seed sample is selected until the total number of samples in a batch is traversed. Each training step takes an action index that selects the corresponding feature from the botnet seed sample for manipulation. The modified sample is tried on the trained $\mathcal{D}$ using the Keras $model.predict$ function, which gives the estimation of the botnet sample being from a minority class. This output estimation is used to set the agent's reward as mentioned in subsection \ref{subsec: reward}. As a result of each step, the reward and the new state (alternatively, the manipulated botnet sample) are returned to the agent. The details of the hyperparameters of the DRL attacker part have been mentioned in Table \ref{table: DRL_hyper} and Table \ref{table: DRL_NN_hyper}.

\begin{table}[tb!]

\centering
\caption{Hyperparameters of DRL Attacker}
\label{table: DRL_hyper}
\centering
\begin{tabu} to 1\textwidth{|m{2.8cm}|m{3cm}|}

\hline
 
  \centering \textbf{Parameter} &
  \centering \textbf{Value} \\ 
  \hline
 
   \centering  Agent Type  & 
      \centering  DQN\\

  \hline
     \centering  Action Space & 
      \centering  13 \\

        \hline
     \centering  Policy & 
      \centering  BoltzmannQPolicy \\

\hline
     \centering  Double DQN & 
      \centering  True \\

      \hline
     \centering  Target Model Update & 
      \centering  1e-3 \\

      \hline
     \centering  Number of turns & 
      \centering  13 \\
  \hline

     \centering Number of rounds &
   \centering 256\\

  \hline

 \end{tabu}
 \end{table}

\begin{table}[tb!]

\centering
\caption{Hyperparameters of DRL Neural Network}
\label{table: DRL_NN_hyper}
\centering
\begin{tabu} to 1\textwidth{|m{2.8cm}|m{3cm}|} 

  \hline

  \centering \textbf{Parameter} &
  \centering \textbf{Value} \\ 
  \hline
 
  \centering  Network Type & 
  \centering FFNN\\

  \hline

  \centering  Number of Layers  &
  \centering 4\\

  \hline

\centering  Activations  &
 \centering ReLU, linear \\
    \hline

  \centering Neurons in input layer  &
  \centering size of observation space\\
\hline

\centering  Neurons in layer 1   &
\centering 64\\
  \hline
  
\centering  Neurons in layer 2   &
\centering 128 \\
  \hline

  \centering  Neurons in output layer   &
  \centering number of actions \\
  \hline

     \centering Layer Regularization &
  \centering $BatchNorm$  \\

  \hline

 \end{tabu}
 \end{table}

\section{Results \& Discussion}
\label{sec: results}
\begin{figure}[tb!]
\centering
\includegraphics[width=8.8cm, height=8.1cm]{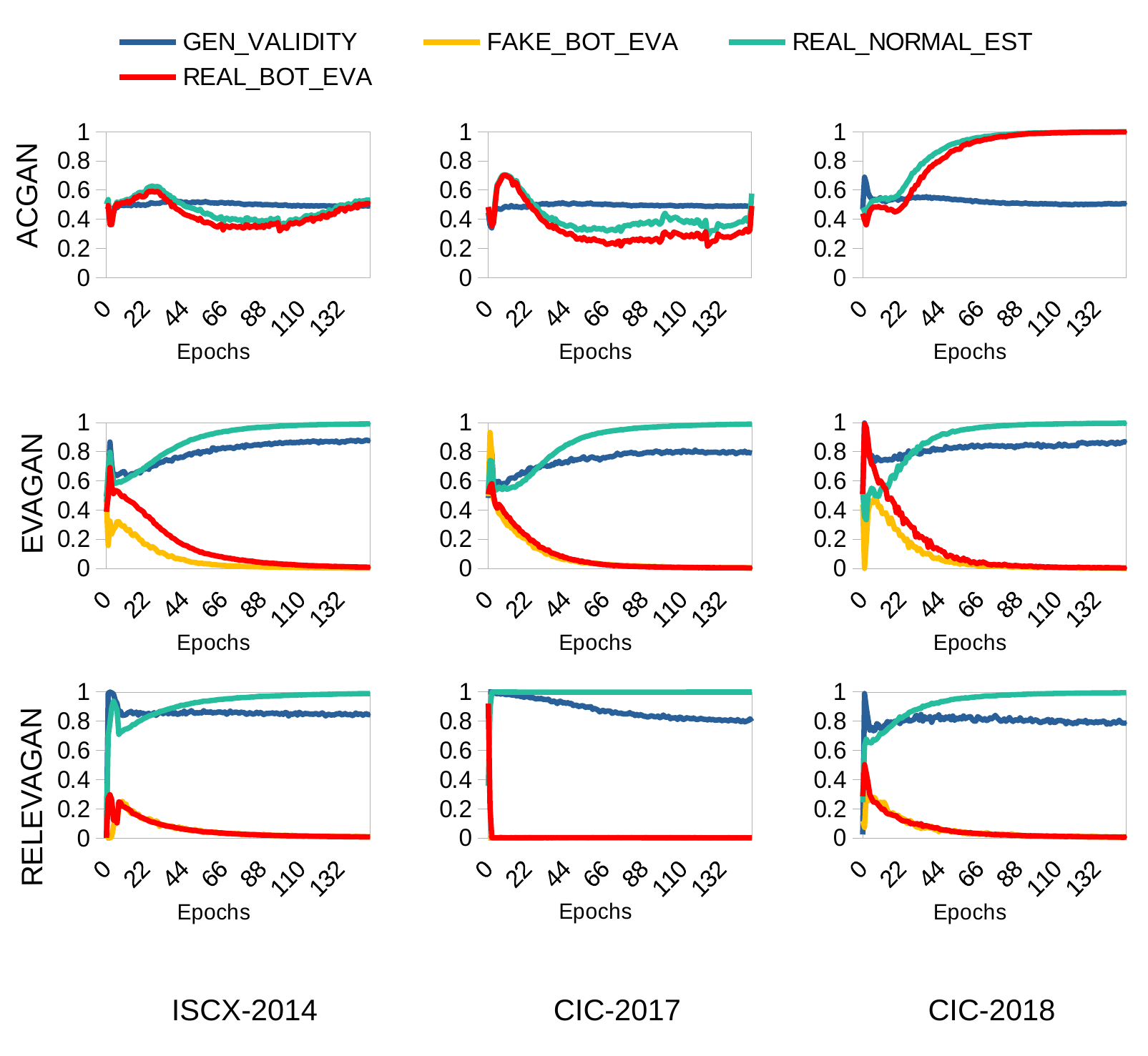}
\caption{The estimations on test data and data generated by the relative GANs generated data}
\label{fig: estimations}
\end{figure}

\begin{figure}[tb!]
\centering
\includegraphics[width=8.8cm, height=8.5cm]{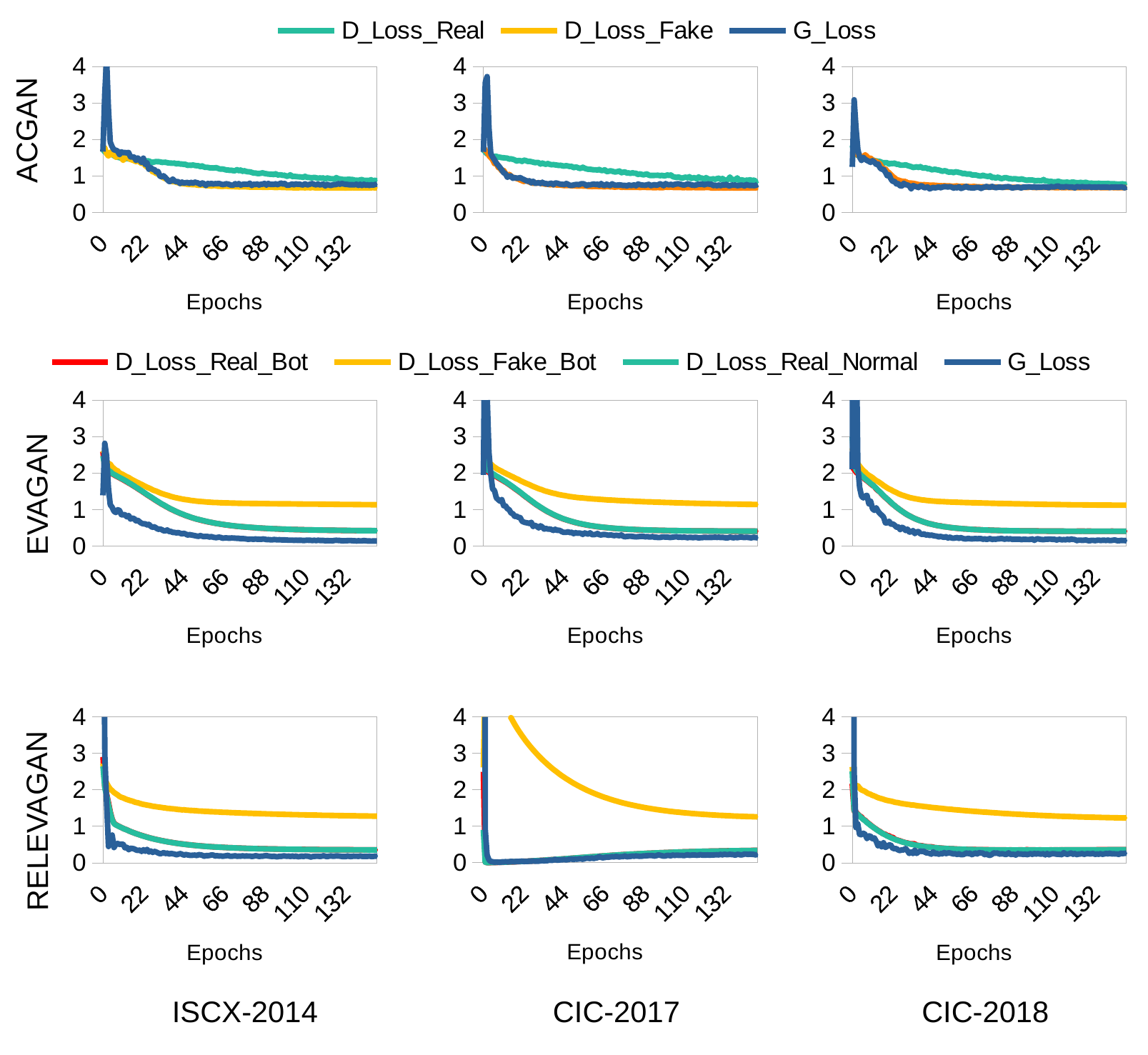}
\caption{The training losses for ACGAN, EVAGAN and RELEVAGAN}
\label{fig: losses}
\end{figure}

The performance analysis of RELEVAGAN is identical to that of EVAGAN for botnet datasets. The metrics used were generated samples validity (GEN\_VALIDITY), fake/generated botnet samples evasion (FAKE\_BOT\_EVA), real normal/majority class estimation (REAL\_NORMAL\_EST), and real botnet/minority class evasion (REAL\_BOT\_EVA). Figure \ref{fig: estimations} shows the results for the estimations of ACGAN, EVAGAN and RELEVAGAN for comparison. The mathematical expressions for the evaluation metrics have been represented using Equations \ref{eq15}-\ref{eq18}. These estimations were computed using the Keras $model.predict$ function. The details of these metrics can be found in EVAGAN paper \cite{randhawa2021evagan}. Figure \ref{fig: losses} illustrates the losses of $\mathcal{D}$ for real and fake minority classes and majority/normal classes and of $\mathcal{G}$ for ACGAN, EVAGAN and RELEVAGAN. 


\begin{equation} \label{eq15}
\begin{aligned}
GEN\_VALIDITY = \frac{\sum [
\hat {\mathcal{G}}(z,c_m)[0] ]}{N}
 \end{aligned}
\end{equation}


\begin{equation} \label{eq16}
\begin{aligned}
FAKE\_BOT\_EVA =\frac{\sum [
\hat {\mathcal{G}}(z,c_m)[1] ]}{N}
 \end{aligned}
\end{equation}


\begin{equation} \label{eq17}
\begin{aligned}
REAL\_NORMAL\_EST = \frac{\sum {[
\hat {\mathcal{X}}_{normal_{test}}}[2] ]}{N}
 \end{aligned}
\end{equation}


\begin{equation} \label{eq18}
\begin{aligned}
REAL\_BOT\_EVA =\frac{\sum {[
\hat {\mathcal{X}}_{botnet_{test}}}[1]]}{N}
 \end{aligned}
\end{equation}

\subsection{Detection Performance}
The discussion on the performance comparison of ACGAN with EVAGAN has been mentioned in the EVAGAN paper; however, in Figure \ref{fig: estimations}, the estimations for ACGAN have also been included for a clear comparison. The performance of ACGAN in discriminating the minority class deteriorates in low data regimes, especially in cybersecurity datasets. On the other hand, $\mathcal{D}$ of EVAGAN discerns the difference between the minority of majority classes. However, regarding RELEVAGAN, the estimations tend toward the desired values quicker than EVAGAN after a few initial unstable values. The RELEVAGAN was pre-trained for the values coming from the real and generated sources, especially from epochs 0-44. This is the contribution of the DRL attacker part. The most interesting case occurs for the CIC-2017 dataset, which took only a couple of epochs to train itself and prepare for both the minority and majority classes. The role of DRL with EVAGAN demonstrates the value addition, especially for the CIC-2017 dataset. This pattern encourages us to further explore the potential of RELEVAGAN on other datasets, which we leave to future work.

We are using the EVAGAN as the base model, which does not require adversarial training of dedicated ML classifiers because the $\mathcal{D}$ of EVAGAN itself works as an adversarial aware botnet detector. So RELEVAGAN also does not need the adversarial training; however, the evasions that the DRL attacker produces need to be fed to the $\mathcal{D}$. This step imitates the back-propagation step in GAN training. 

GEN\_VALIDITY also illustrates the early convergence of $\mathcal{G}$ which achieves the Nash equilibrium quickly in RELEVAGAN while the $\mathcal{G}$ of EVAGAN is still learning. Since our main objective is to improve the detection performance of our model for botnet in low data regimes, we do not need to let the $\mathcal{G}$ train for a larger number of epochs. For example, if we achieve 100\% detection performance, as in the case of the CIC-2017 dataset, we can stop the GAN training after a couple of steps.

\subsection{Stability}
Figure \ref{fig: losses} demonstrates the $\mathcal{D}$ and $\mathcal{G}$ losses for ACGAN, EVAGAN and RELEVAGAN. It turns out that the losses tend to converge for all the GANs. The values for RELEVAGAN tend to achieve the lowest point sooner than both EVAGAN and ACGAN. The values for D\_Loss\_Fake for RELEVAGAN are higher than other GANs. This can be because the $\mathcal{D}$ of RELEVAGAN is struggling to discriminate between the evasion generated by the DRL and the $\mathcal{G}$. The evasion samples coming from DRL are labelled as 'REAL', and when similar samples arrive from the $\mathcal{G}$, the wrong estimation increases the loss. However, the overall detection performance improves due to the reason that the $\mathcal{G}$ now tends to generate the samples within the semantic constraints. There can be a possibility of mode collapse here; however, the detection performance is better than EVAGAN, so we can disregard that factor.

\subsection{DRL Attacker Reward/Evasions}
\begin{figure}[tb!]
\centering
\includegraphics[width=5.71cm, height=4cm]{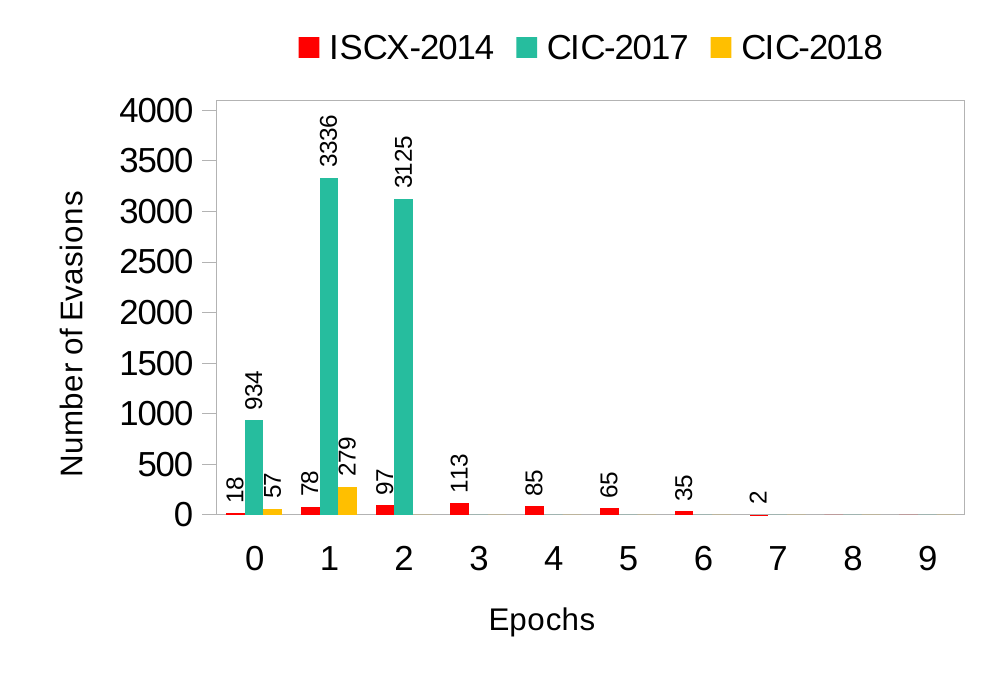}
\caption{RELEVAGAN reward/evasions during DRL attack on EVAGAN discriminator for the three datasets}
\label{fig: evasions}
\end{figure}

Figure \ref{fig: evasions} shows the number of evasions the DRL attacker generated or the reward collected during the exploration or training phase of RELEVAGAN of the first ten epochs. No evasions happen after epoch number eight. Note that the highest number of evasions were in the case of the CIC-2017 dataset, which gives the best results for estimations in Figure \ref{fig: estimations}. A similar pattern is seen in the cases of ISCX-2014 and CIC-2018 datasets, but no evasions for the CIC-2018 dataset were generated after epoch three. If we correlate the number of generated evasions by the DRL attacker with the performance of the $\mathcal{D}$, it turns out that there is an inverse relationship between the number of evasions and convergence of the training of RELEVAGAN. Trying the model on other datasets can give more insights into this relationship.

\subsection{Time Complexity}
\begin{figure}[tb!]
\centering
\includegraphics[width=5.71cm, height=4cm]{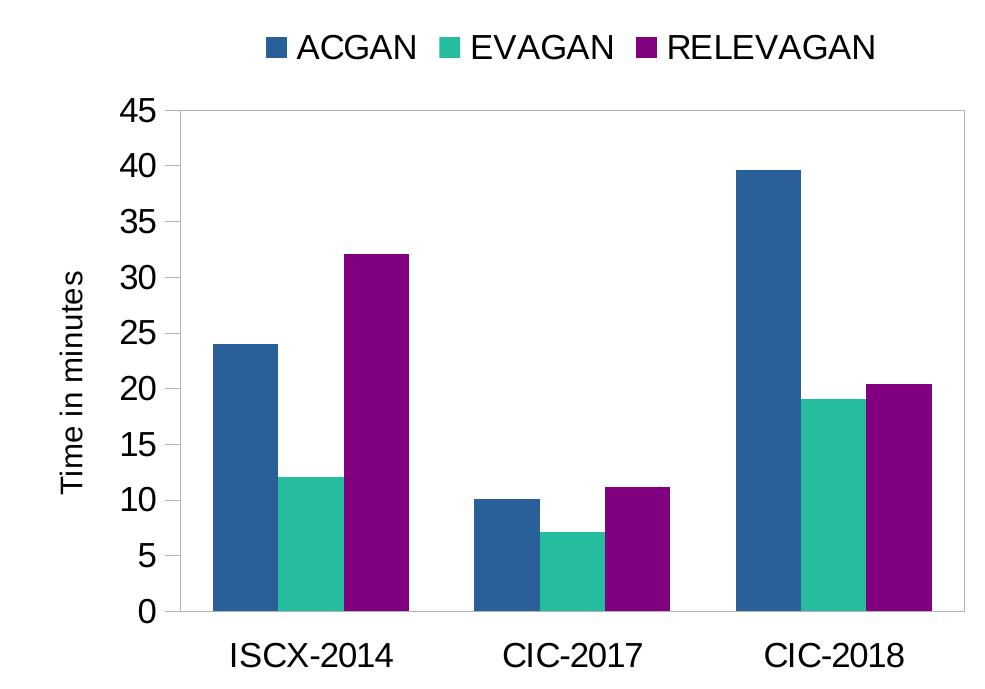}
\caption{Time complexity}
\label{fig: timeComplexity}
\end{figure}

Figure \ref{fig: timeComplexity} shows the training time complexity of ACGAN, EVAGAN and RELEVAGAN for the three datasets. RELEVAGAN training time for 150 epochs is always greater than that of EVAGAN because DRL has its own cost. However, we get the benefit of early convergence to achieve the maximum detection performance as manifested in Figure \ref{fig: estimations} for the CIC-2017 dataset. Hence we achieve the performance of 100\% in the time way less than taken by EVAGAN, but it turns out to depend on the dataset. As for the case of the ISCX-2014 dataset, the time complexity is even more than ACGAN's training time. Hence, we consider working towards more datasets for greater insight into the time complexity pattern as well.

\section{Conclusion}
\label{sec: conclusion}
The semantic aware adversarial botnet detector is essential for countering modern evasion attacks. In this regard, the detection models must proactively be aware of the possible adversarial perturbations. Researchers employ deep reinforcement learning to generate adversarial evasion examples to preserve the original functionality of the botnet/malware samples. The motivation is to generate samples that could be used later for adversarial training of the ML classifiers. We propose RELEVAGAN, which proves to be an adversarial semantic-aware evasion detection model that does not need exclusive adversarial training. The discriminator. RELEVAGAN is based on EVAGAN, which did not consider the semantic awareness that RELEVAGAN resolves. We have used the three datasets used by the EVAGAN paper for a better comparison and coherency. The results demonstrate the supremacy of RELEVAGAN to EVAGAN in early convergence to the detection model training. 

We highly recommend testing RELEVAGAN's performance against other cybersecurity datasets for future research. Few-shot learning could also be tried on the datasets used in this paper to compare performance.



\bibliographystyle{IEEEtran}
\bibliography{ref}

\begin{IEEEbiography}[{\includegraphics[width=1in,height=1.25in,clip,keepaspectratio]{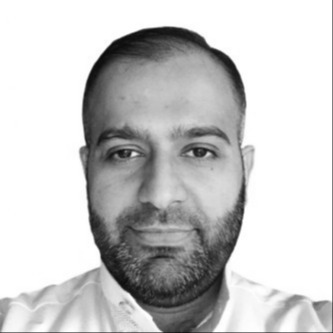}}]{Rizwan Hamid Randhawa} received a BS degree in Electronic Engineering from International Islamic University Islamabad, Pakistan and Master in Computer Science from Information Technology University, Lahore, Pakistan. He has vast experience with embedded systems in Pakistan's private and public sector organisations. He is pursuing his PhD in Computer Science from Northumbria University, Newcastle upon Tyne, UK. His research interests include AI-based botnet detection, IoT Security and Embedded Systems Design \& Development for IoT platforms.

\end{IEEEbiography}

\begin{IEEEbiography}[{\includegraphics[width=1in,height=1.25in,clip,keepaspectratio]{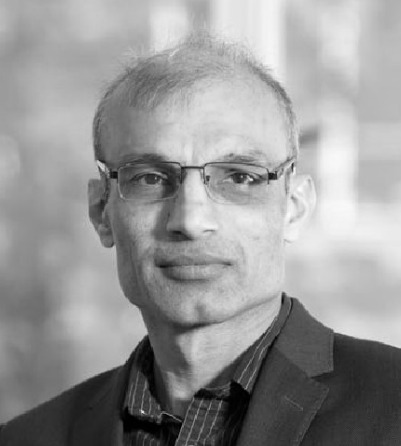}}]{Nauman Aslam} is a Professor in the Department of Computer and Information Science, Northumbria University, UK. Before joining Northumbria University as a Senior Lecturer in 2011, he worked as an Assistant Professor at Dalhousie University, Canada. He received his PhD in Engineering Mathematics from Dalhousie University, Canada, in 2008. Dr Nauman leads the Network Systems and Security research group at Northumbria University. His research interests cover diverse but interconnected areas related to communication networks. His current research focuses on addressing wireless body area networks and IoT, network security, QoS-aware communication in industrial wireless sensor networks, and Artificial Intelligence (AI) application in communication networks. He has published over 100 papers in peer-reviewed journals and conferences. Dr Nauman is a member of IEEE.
\end{IEEEbiography}

\begin{IEEEbiography}[{\includegraphics[width=1in,height=1.25in,clip,keepaspectratio]{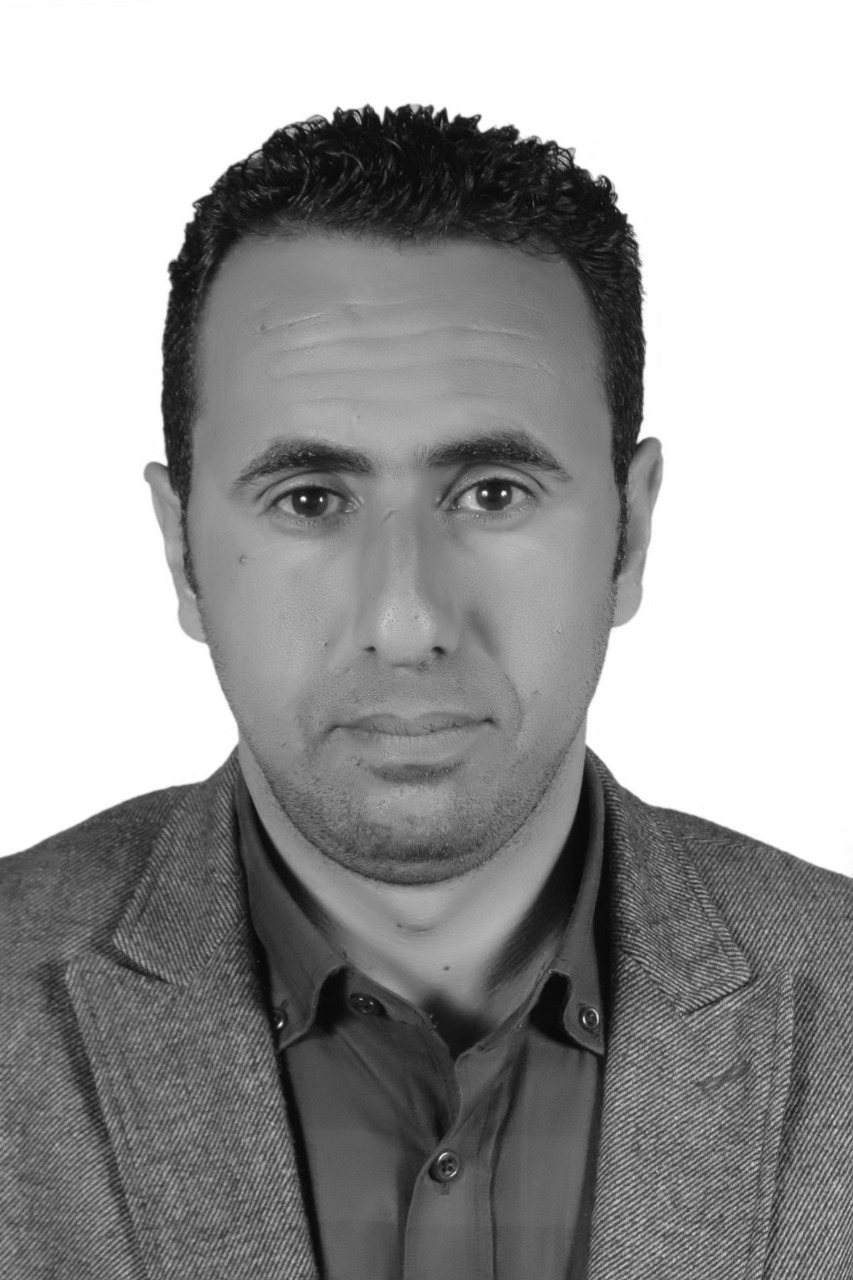}}]{Mohammed Alauthman} received his PhD degree from Northumbria University Newcastle, the UK, in 2016. He received a B.Sc. degree in Computer Science from Hashemite University, Jordan, in 2002, and an M.Sc. degree in Computer Science from Amman Arab University, Jordan, in 2004. Currently, he is Assistant Professor at the Information Security Department, Petra University, Jordan. His research interests include Cyber-Security, Cyber Forensics, Advanced Machine Learning and Data Science applications.
\end{IEEEbiography}

\begin{IEEEbiography}[{\includegraphics[width=1in,height=1.25in,clip,keepaspectratio]{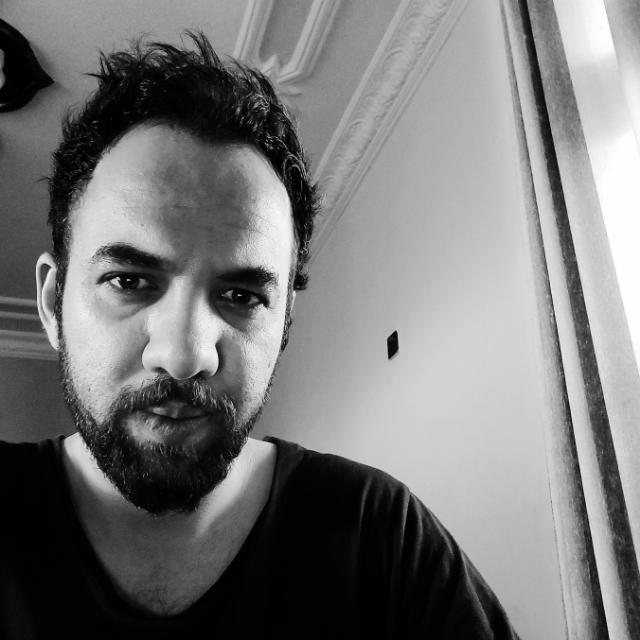}}]{Muhammad Khalid} received the PhD degree in computer science from Northumbria University, Newcastle Upon Tyne, U.K. He is currently working as lecturer at School of Computer Science, University of Hull, UK. Earlier he worked as a research fellow at University of Lincoln, UK. His research interests include reinforcement learning, autonomous systems, safety in robotics, Internet of Things, and autonomous valet parking. 
\end{IEEEbiography}

\begin{IEEEbiography}[{\includegraphics[width=1in,height=1.25in,clip,keepaspectratio]{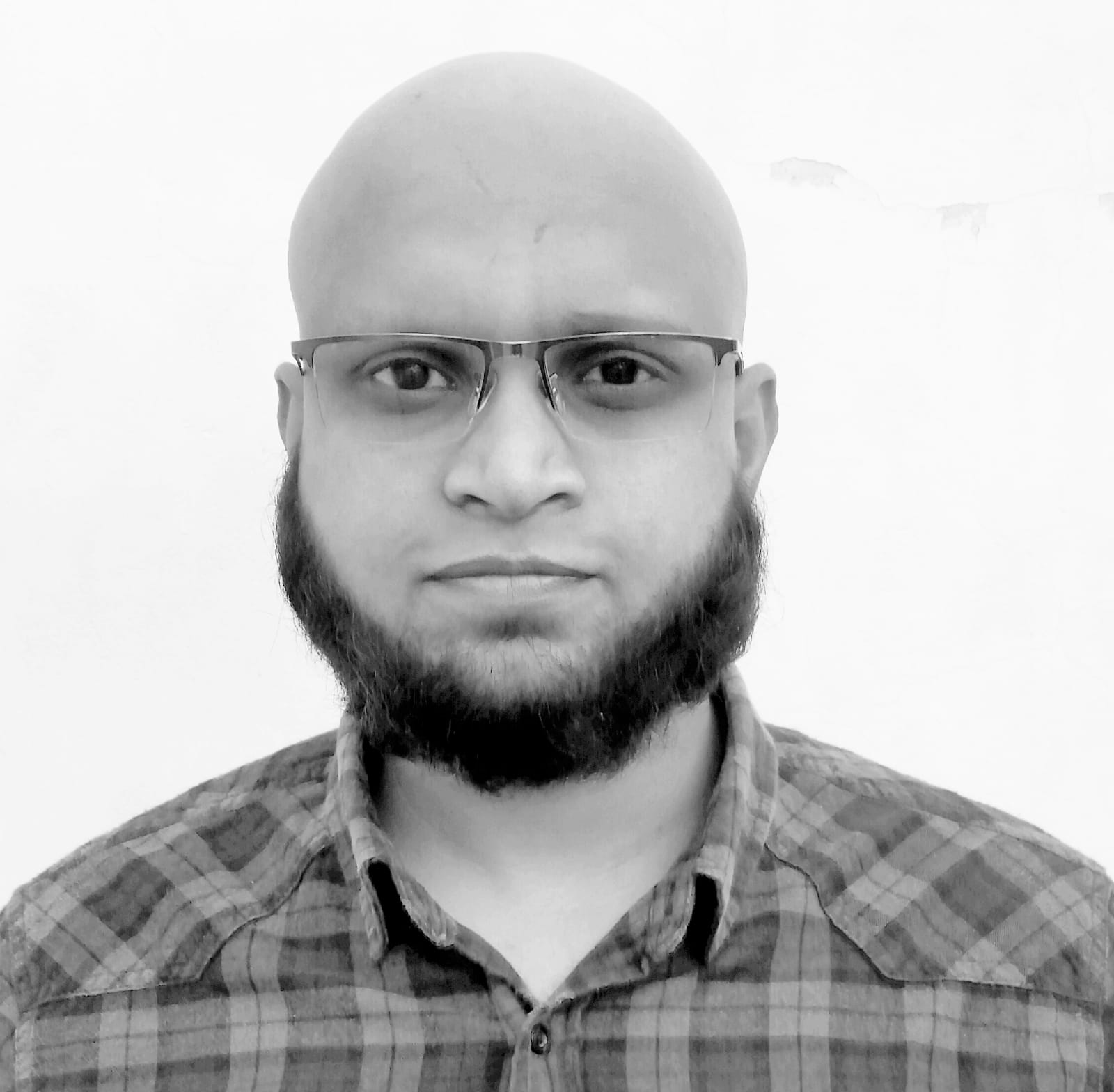}}]{Husnain Rafiq} received the B.S. and M.S. degrees in Computer Science from the Capital University of Science and Technology, Islamabad, Pakistan in 2015 and 2017, respectively. From 2015 to 2018, he was a Jr. Lecturer with the Capital University of Science and Technology, Islamabad, Pakistan. He is pursuing a PhD from Northumbria University Newcastle upon Tyne, UK. His area of research includes Information Security and Forensics, Machine learning and Malware Analysis. 
\end{IEEEbiography}

\end{document}